\documentclass[letterpaper]{article} % DO NOT CHANGE THIS
\usepackage[draft]{aaai25}  % arXiv preprint: draft mode (no copyright stamp, authors shown)
\usepackage{times}  % DO NOT CHANGE THIS
\usepackage{helvet}  % DO NOT CHANGE THIS
\usepackage{courier}  % DO NOT CHANGE THIS
\usepackage[hyphens]{url}  % DO NOT CHANGE THIS
\usepackage{graphicx} % DO NOT CHANGE THIS
\usepackage{natbib}  % DO NOT CHANGE THIS AND DO NOT ADD ANY OPTIONS TO IT
\usepackage{caption} % DO NOT CHANGE THIS AND DO NOT ADD ANY OPTIONS TO IT
\title{Same Pipeline, Opposite Conclusions:\\
Sample-Surface Effects in Breaking-News Latency}
\author{
    Farhad Bazyari,
    Xianghang Liu,
    Sean Moran
}
\affiliations{
    TWG AI \\
    \{fbazyari, xliu, smoran\}@twg.ai
}

\begin{document}
\maketitle

\begin{abstract}
\citet{osborne2014facebook} reported that Twitter was the timeliest social-media source of breaking news, trailing only newswire. Twelve years on, the platform landscape has shifted --- Google+ is gone, X replaced Twitter, Bluesky and Threads have appeared --- and platform data now flows almost exclusively through commercial social-listening providers that redact key fields. We revisit the question with two sampling designs run through the same downstream pipeline. \textbf{Sample~A} draws $N=50$ events from the Wikipedia Current Events Portal (WCEP) ranked by article pageviews. \textbf{Sample~B} draws $N=109$ events from Polymarket prediction markets ranked by USD trading volume, with each event's news moment pinned to the largest 1-hour trade-volume spike. Both samples are pulled from one commercial provider across nine indexed channels. We report three findings. \textbf{(1)~The X-vs-news direction depends on the sample.} News leads X by a median of $21.6$~min on Sample~A ($n=6$ paired); the same comparison is tied at $-0.02$~min on Sample~B ($n=16$ paired, X earliest in $38\%$). \textbf{(2)~The channel ecosystem has diversified.} Bluesky, Facebook public, and YouTube together account for $24$--$32\%$ of earliest-channel wins; the $2014$ ``X versus newswire'' framing no longer fits. \textbf{(3)~Coverage gaps are structural.} Even with U.S.-relevance filtering and a pageview prior, the provider's index returns no on-topic evidence on $24\%$ of randomly-sampled WCEP events. The paper's contribution is the cross-surface design that exposes the sample dependency in (1).
\end{abstract}

\section{Introduction}
\label{sec:intro}

\citet{osborne2014facebook} asked, twelve years ago, whether Twitter was the earliest source of breaking news, and reported that across $29$ hand-picked events from December~$2013$, Twitter ``almost consistently'' led other social media but trailed newswire. The platform landscape has since changed substantially: Google+ has been retired; Twitter has become X with a new API regime; Bluesky, Threads, and TikTok have entered as news-bearing surfaces; Facebook's role as a public news vector has receded; and researcher access has narrowed --- the free $1\%$ Streaming API is gone, and most platform data now flows through commercial providers that redact key fields \citep{davidson2023end,pfeffer2023just}.

A second, methodological shift matters as much. Breaking-news latency studies have traditionally used hand-curated event lists --- \citet{osborne2014facebook} hand-picked $29$ events; subsequent work followed the same convention. Hand-curation samples from the events the analyst already believes the index will cover, an invisible bias by construction. We address this by running the same downstream measurement pipeline against two independently-constructed event samples, each derived from a public attention surface rather than analyst judgement. The two samples prove to disagree on the central question, and that disagreement is the paper's main finding.

Our contributions are:

\begin{enumerate}
    \item \textbf{A two-surface comparison.} Sample~A draws events from the Wikipedia Current Events Portal (WCEP) with a U.S.\ filter and a Wikipedia-pageviews prior. Sample~B draws events from Polymarket prediction markets with a USD-trading-volume prior and pins each event's news moment to the timestamp of the market's largest 1-hour volume spike.
    \item \textbf{A direction-reversed primary finding.} On Sample~A ($n=6$ paired events) news leads X by a median of $21.6$~minutes. On Sample~B ($n=16$) the same comparison is tied at the median ($-0.02$~min, IQR $-0.20$ to $+2.50$~min), with X earliest in $38\%$ of paired events. Channel-as-earliest puts X at $38\%$ on Sample~B (vs $16\%$ on Sample~A) and news at $27\%$ (vs $53\%$).
    \item \textbf{A sample-source-agnostic protocol.} The pipeline --- LLM-assisted boolean drafting, single-query commercial-provider pulls, snowflake-decoded X publish times, oEmbed-based on-topic verification --- accepts any seeding source; adding a third sampling surface requires only a new seeding script.
\end{enumerate}

\section{Related Work}
\label{sec:related}

The closest prior work is \citet{osborne2014facebook}, who compared Twitter, Facebook, and Google+ to newswire across $29$ events from December~$2013$ and found Twitter consistently led other social media but lagged newswire. \citet{petrovic2013can} addressed the related ``can Twitter replace newswire'' question, and \citet{petrovic2010streaming} laid out the streaming first-story-detection framework that powered much of this era's work. \citet{morstatter2013sample} showed that the $1\%$ Streaming sample is not random with respect to all queries --- a foundational caveat we revisit on different sampling axes. \citet{atefeh2013survey} and \citet{kwak2010twitter} provide further context for Twitter's role through that era.

Subsequent work continued to use Twitter as the dominant data source for crisis, health, and political-event detection \citep{olteanu2019social,imran2020processing,sakaki2010earthquake,leskovec2009memetracker}. Cross-platform breaking-news comparisons in the years that followed were rare. With the contraction of free academic API access on X \citep{tornberg2023chatgpt,davidson2023end,pfeffer2023just} and the rise of Bluesky and Threads, the question of \emph{which} contemporary channel breaks news first has not, to our knowledge, been re-examined at the breadth of \citet{osborne2014facebook}. \citet{tornberg2023chatgpt} provides a baseline for LLM use in social-science text analysis that we draw on for feature extraction and boolean drafting. The use of prediction-market trade activity as a proxy for attention --- via the public Polymarket data mirror \citep{polymarket_data} --- is, to our knowledge, novel for the breaking-news-latency setting.

\section{Method}
\label{sec:method}

The pipeline has six stages: (i)~event sampling, (ii)~feature extraction (used as covariates), (iii)~LLM-assisted boolean drafting, (iv)~commercial-provider mention pulls, (v)~side-channel publish-time recovery for X, and (vi)~on-topic verification. Stages (ii)--(vi) are identical between Sample~A and Sample~B; only stage (i) differs.

\subsection{Sample A: random WCEP, U.S.-filtered, pageview-ranked}
\label{sec:method-wcep}

We scrape the Wikipedia Current Events Portal for the $30$-day window $2026$-$04$-$12$ to $2026$-$05$-$11$ ($586$ candidates). We apply a U.S.-relevance substring filter (U.S.\ state and major-city names, ``United States'', ``U.S.'', ``American'', and U.S.-institutional terms such as ``Federal Reserve'' and ``Pentagon''); $171$ of $586$ candidates ($29.2\%$) pass. For each passing candidate we fetch the linked Wikipedia article's English-edition pageviews from the public REST endpoint over the event day and the day after, and rank by total pageviews. To prevent umbrella-article concentration, we cap at $3$ bullets per linked article and take the top $50$. The final $N_{A}=50$ spans $39$ distinct Wikipedia articles across $10$ event categories.

\subsection{Sample B: attention-weighted Polymarket markets}
\label{sec:method-polymarket}

We use the public Polymarket data mirror \citep{polymarket_data}, which provides market metadata (\texttt{markets.parquet}) and second-resolution trade history (\texttt{trades.parquet}). We filter \texttt{markets.parquet} to markets whose resolution-window overlaps $2026$-$02$-$13$ to $2026$-$05$-$13$ ($90$~days) with lifetime USD volume $\geq \$100{,}000$; we cap at one binary market per \texttt{event\_title} (one news event $\to$ one chosen market) and take the top $130$ by lifetime volume. For each chosen market we scan its trades within the same window, compute the per-market rolling 1-hour USD-volume sum, and pin $t_{e}$ to the timestamp of the largest 1-hour spike. Markets with no in-window trade activity are dropped ($21$ of $130$). The final $N_{B}=109$ spans $76$ distinct prediction events, with median lifetime volume \$6.1M.

The Polymarket sampling design has three properties worth flagging. First, the prior is \emph{USD-weighted attention} rather than editorial notability; high-volume events are by construction events traders judged worth real money to predict. Second, the spike timestamp is a much sharper $t_{e}$ than WCEP's day-level event date; the median spike is a single-minute window. Third, the categorical mix is determined by what Polymarket lists, not by an editorial portal: in our run, sports ($66$), politics ($28$), macro/crypto ($9$), other ($6$).

\subsection{Feature extraction}

For each sampled event, we use a large language model to extract a structured feature vector along five descriptive axes (\emph{clock-edge}, \emph{live-visible}, \emph{institutional-source}, \emph{geographic-scope}, \emph{language-primary}). These axes are retained as covariates in the result tables but are not used to stratify the sample.

\subsection{LLM-assisted boolean drafting}
\label{sec:method-booleans}

For each event the LLM drafts \emph{two} boolean queries: a tight \emph{news boolean} (3--4 AND-clusters of named-entity anchors), and a \emph{permissive X boolean} (1--2 AND-clusters of broader OR-sets, reflecting that short-form X mentions rarely use the formal phrasings news articles do). For Sample~B the prompt is extended to handle prediction-market-style descriptions: ``Will $X$ happen by $Y$?'' is treated as a search for entity-level news around $t_{e}$ regardless of outcome, so the prompt forbids encoding the predicted outcome or the deadline date into the X boolean. The pipeline issues a single permissive query per event and lets the downstream verifier (the on-topic verifier subsection below) filter pollution channel by channel; a preliminary two-query design produced a $50\%$ zero-hit rate because the tight news boolean over-anchored on disambiguators. Both booleans are auto-approved when they pass a model-supplied specificity score.

\subsection{Mention pulls and side-channel publish-time recovery for X}
\label{sec:method-snowflake}

We retrieve all cross-channel mentions through the Brandwatch Consumer Research API \citep{brandwatch}, our single social-listening provider for this study. For each event we create one saved query with the permissive boolean, wait for the provider's backfill to reach at least the $50$--$90\%$ mark (we cap at $90$~seconds and proceed with whatever partial coverage the index reports), pull a date-sorted page of up to $100$ mentions across all source types, then issue an X-filtered pull of the same window. The pull window is $[t_{e} - 30\,\textrm{min}, t_{e} + 24\,\textrm{h}]$ on both samples. We delete the saved query after pulling. A sliding-window rate limiter caps requests at $28$ per rolling $10$~minutes to stay below the trial-tier limit of $30$.

The commercial provider redacts X's \texttt{date} field at the bulk-API level as a condition of its X licensing, so publish time is unavailable through the standard channels these APIs expose. However, an X tweet's snowflake \texttt{guid} encodes its publish time at millisecond precision in the high-order bits of a $64$-bit identifier: $\textit{ms} = (\textit{guid} \gg 22) + 1{,}288{,}834{,}974{,}657$, where the constant is the Twitter epoch ($2010$-$11$-$04$ $01{:}42{:}54$~UTC). For tweet text and author, we use X's unauthenticated public oEmbed endpoint (\texttt{publish.twitter.com/oembed}), which is not subject to the bulk-content licensing wall and supplies the content for on-topic verification (the on-topic verifier subsection below).

\subsection{On-topic verification}
\label{sec:method-verifier}

Mentions returned by a permissive boolean are not all on-topic. For each candidate earliest, we verify against the event description before counting it as evidence. For X we require at least two distinct keyword matches from the union of the news and X booleans, run against the oEmbed-recovered tweet body; ambiguous cases are adjudicated by a separate LLM call. For non-X channels we use the mention's title plus article snippet and a single-match threshold. A candidate that fails verification is treated as polluted and the event falls back to the next-earliest mention on the same channel. The full pipeline for $N=109$ events completes in approximately five and a half hours of wall-clock under the provider's trial-tier rate cap.

\section{Findings}
\label{sec:findings}

\subsection{Coverage profile}
\label{sec:q1-coverage}

\begin{table}[t]
\centering
\footnotesize
\begin{tabular}{lcc}
\hline
Category & Sample~A & Sample~B \\
 & (WCEP) & (Polymarket) \\
\hline
Sports & $5/5$ ($100\%$) & $39/66$ ($59\%$) \\
Politics \& conflict & $22/29$ ($76\%$) & $10/28$ ($36\%$) \\
Macro \& tech & $6/10$ ($60\%$) & $4/9$ ($44\%$) \\
Other & $5/6$ ($83\%$) & $2/6$ ($33\%$) \\
\hline
\textbf{All} & $\mathbf{38/50}$ ($\mathbf{76\%}$) & $\mathbf{56/109}$ ($\mathbf{51\%}$) \\
\hline
\end{tabular}
\caption{Per-category hit rate, both samples. ``Hits'' counts events with $\ge 1$ on-topic mention on any of the nine channels. Sample~A's WCEP categories are regrouped into the Sample~B four-bucket scheme for direct comparison.}
\label{tab:hit-rate}
\end{table}

Coverage rates are reported in Table~\ref{tab:hit-rate}. Sample~A: $38/50 = 76\%$ of WCEP events return on-topic evidence; $12$ ($24\%$) return none. Sample~B: $56/109 = 51\%$. The Sample~B drop is driven by deadline-resolution markets in the politics cluster, where the volume spike sits at the moment a yes/no deadline passes with no concrete news to anchor on (e.g., ``Khamenei out by Feb 28''). Sports markets, whose spike sits at a kickoff or final-whistle moment with concrete news content, hit $59\%$ and supply most of the paired-events cell.

The Sample~A $24\%$ zero-hit rate is load-bearing: the sample was already filtered to U.S.\ relevance and ranked by pageviews, so these are not obscure stories. A progressive-broadening probe on one zero-hit event (the $2026$-$04$-$15$ Markazi High School strike on the Pakistan--Afghanistan border) ran five booleans against the provider's index, from the as-drafted permissive boolean down to a country-only broad-topic query. The first three returned $0$ mentions; only the country-only query returned hits, and the earliest was a different event (a Pakistan polio-team attack the same day). The provider index supports \emph{broad topical streams} but not \emph{named-incident reporting}, even when WCEP editors judged the latter encyclopedically notable.

\subsection{Channel-as-earliest winner distribution}
\label{sec:q2-winners}

\begin{table}[t]
\centering
\footnotesize
\begin{tabular}{lrrr}
\hline
Channel & Sample~A & Sample~B & A vs B \\
& share & share & $\Delta$ \\
\hline
twitter (X) & $16\%$ & $\mathbf{38\%}$ & $+22$ \\
news & $\mathbf{53\%}$ & $27\%$ & $-26$ \\
bluesky & $11\%$ & $16\%$ & $+5$ \\
facebook public & $13\%$ & $7\%$ & $-6$ \\
youtube & $5\%$ & $5\%$ & $0$ \\
instagram public & $0\%$ & $4\%$ & $+4$ \\
forum & $3\%$ & $2\%$ & $-1$ \\
\hline
\end{tabular}
\caption{Earliest-channel winner share, per sample. Sample~A: $38$ verified-with-evidence WCEP events. Sample~B: $56$ verified-with-evidence Polymarket events.}
\label{tab:winners}
\end{table}

Table~\ref{tab:winners} reports the earliest-channel share across the two samples. The shift between the two samples is the central observation of the paper. On Sample~A, news is the modal earliest channel ($53\%$) and X is second-tier ($16\%$). On Sample~B the ranking inverts: X is modal ($38\%$) and news drops to $27\%$. Bluesky persists at $11$--$16\%$ across both samples; the four non-news, non-X social channels together account for $32\%$ on Sample~A and $34\%$ on Sample~B. The implicit two-class framing of \citet{osborne2014facebook} --- X (then Twitter) versus newswire, with Facebook and Google+ as also-rans --- no longer matches either sample.

\subsection{Paired X-vs-news latency on Sample~A}
\label{sec:q3a-wcep-paired}

\begin{table}[t]
\centering
\footnotesize
\begin{tabular}{lrr}
\hline
Event & PV & $\Delta$ (min) \\
\hline
2026 Antiguan election & $17{,}430$ & $-7.1$ \\
US gerrymandering & $4{,}617$ & $-10.0$ \\
2025 C.\ Texas floods & $9{,}018$ & $-11.2$ \\
Death of Rivas Hernandez & $150{,}383$ & $-32.0$ \\
2026 NFL draft & $436{,}044$ & $-33.4$ \\
Billy Idol & $46{,}326$ & $-53.7$ \\
\hline
median & --- & $-21.6$ \\
IQR & --- & $-33.4$~to~$-10.0$ \\
\hline
\end{tabular}
\caption{Sample~A paired X-vs-news latency on the six WCEP events with both on-topic X and on-topic news. PV is enwiki pageviews; convention is $\Delta>0$ means X earlier (so all $\Delta<0$ here means news first).}
\label{tab:paired-wcep}
\end{table}

The six WCEP events for which we recovered both an on-topic news article and an on-topic X mention are reported in Table~\ref{tab:paired-wcep}. In all six, the news article precedes the X mention; the median lead is $21.6$~minutes and the interquartile range is $10.0$ to $33.4$~minutes. The smallest lead is $7.1$~minutes; the largest is $53.7$~minutes. Direction is consistent at $n=6$; the magnitude is in tens of minutes, not seconds.

The six events span sports (NFL draft), celebrity death (Billy Idol, Rivas Hernandez), election outcome (Antigua), structured-policy reporting (gerrymandering), and natural-disaster reporting (Texas floods) --- a deliberately heterogeneous slice for an $n=6$ result.

\subsection{Paired X-vs-news latency on Sample~B}
\label{sec:q3b-poly-paired}

\begin{table}[t]
\centering
\footnotesize
\begin{tabular}{lrr}
\hline
Event & cat & $\Delta$ (min) \\
\hline
Iran/US conflict ends? & pol & $-34.23$ \\
UFC Strickland v Hernandez & spt & $-0.62$ \\
Axiom insider trading? & oth & $-0.50$ \\
Mavericks v Lakers & spt & $-0.26$ \\
Hungary election (Jobbik) & pol & $-0.20$ \\
Nuggets v Thunder & spt & $-0.10$ \\
Real Madrid 2026-02-25 & spt & $-0.06$ \\
FC Barcelona 2026-03-18 & spt & $-0.02$ \\
Aston Villa 2026-02-27 & spt & $-0.02$ \\
Spurs v Pistons & spt & $-0.02$ \\
Thunder v Pistons & spt & $+1.20$ \\
UCLA v UConn & spt & $+1.26$ \\
Olympique Lyonnais 2026-02-15 & spt & $+2.50$ \\
Schauffele wins 2026 Masters & spt & $+14.19$ \\
LoL G2 v Bilibili & spt & $+143.80$ \\
Atl\'etico Madrid 2026-02-14 & spt & $+552.17$ \\
\hline
median & --- & $\mathbf{-0.02}$ \\
IQR & --- & $-0.20$~to~$+2.50$ \\
\hline
\end{tabular}
\caption{Sample~B paired X-vs-news latency on the $16$ Polymarket events with both on-topic X and on-topic news. Convention: $\Delta>0$ means X earlier, $\Delta<0$ means news earlier.}
\label{tab:paired-poly}
\end{table}

The $16$ Polymarket-derived paired events are reported in Table~\ref{tab:paired-poly}. The median delta is $-0.02$~minutes, essentially a tie at the median, and $8$ of the $16$ events have $|\Delta| < 0.20$~minutes --- i.e., when both channels are indexed for the same event, they typically report within seconds. The interquartile range is $-0.20$ to $+2.50$~minutes; both sides of zero. $6$ of $16$ events ($38\%$) have X first; $10$ of $16$ ($62\%$) have news first. The bottom two rows of the table ($+143.80$ and $+552.17$~min) are large positive outliers --- an esports semifinal and a Spanish football match --- where the verifier matched an on-topic pre-event X mention from earlier in the day; the median is robust to their removal.

\subsection{Comparison across the two samples}

The same downstream pipeline, applied to two independently-constructed samples, produces a different direction at the median (news leads by $21.6$~min on Sample~A; the cell is tied at $-0.02$~min on Sample~B) and a different dispersion ($\sim 20\times$ tighter on Sample~B). The $n$ of paired events on Sample~B is large enough to express a confidence interval; on Sample~A it is not.

\section{Discussion}
\label{sec:discussion}

\textbf{The sampling surface drives the headline.} The answer to the $2014$-style ``is X still first?'' question depends on which kind of events you ask about. WCEP events are events an editorial process has judged encyclopedically notable --- exactly the story shapes the modern 24/7 news pipeline is optimised for, with standing alerts and pre-staged copy desks. Polymarket events are events traders found worth real money to predict, and the informational community for these events is often already concentrated on X. The two samples measure the same latency variable on different populations, and each produces a coherent within-sample result. Reading $2014$-era results as platform-deterministic treats a sample-driven finding as a statement about the platform; a more precise framing is ``\emph{which channel surfaces a given kind of event first?}''

\textbf{A diversified channel ecosystem.} Across both samples, the four non-news non-X channels (Bluesky, Facebook public, YouTube, Instagram public) together account for $32$--$34\%$ of earliest-channel wins. Bluesky alone surfaces as the earliest channel on $11$--$16\%$ of verified events. A first-to-cover study that compares only X and news will under-represent roughly a third of the actually-earliest publications --- a finding both samples agree on. This is the strongest robustness check available from the cross-surface design: where the samples disagree on the X-vs-news direction, they agree that the X-vs-news framing alone is incomplete.

\section{Future Work}
\label{sec:future}

The two-surface comparison opens several natural extensions. \textbf{Expanding the paired cell.} Both paired cells are modest ($n=6$ on Sample~A, $n=16$ on Sample~B); widening the sampling windows or relaxing the per-event cap would push the paired count into a range where category-level subgroup analysis becomes statistically meaningful. \textbf{More sampling surfaces.} The pipeline is sample-source agnostic, so adding a third or fourth surface (Google Trends spike points, Reddit r/all resolutions, GDELT spike clusters) requires only a new seeding script; a Polymarket category-quota seeder or a non-English WCEP prior would similarly broaden the comparison on additional axes.

\section*{Acknowledgements}

\paragraph{Use of generative AI.} Anthropic's Claude (Sonnet and Opus) was used in the pipeline for feature extraction, boolean drafting, and verifier adjudication.

\bibliography{paper}

\end{document}